\documentclass[a4paper,onecolumn,11pt,unpublished]{quantumarticle}
\pdfoutput=1

\usepackage{float}
\usepackage[utf8]{inputenc}
\usepackage[english]{babel}
\usepackage[T1]{fontenc}
\usepackage{amsmath}
\usepackage{amssymb}
\usepackage{amsthm}
\usepackage{hyperref}
\usepackage{tikz}
\usepackage{booktabs}
\usepackage{algorithm}
\usepackage{algpseudocode}
\usepackage{cite}

\newtheorem{theorem}{Theorem}
\newtheorem{lemma}{Lemma}
\newtheorem{proposition}{Proposition}
\newtheorem*{proposition*}{Proposition}
\newtheorem{corollary}{Corollary}
\newtheorem{assumption}{Assumption}
\theoremstyle{remark}

\newcommand{\norm}[1]{\left\lVert #1 \right\rVert}
\newcommand{\bs}[1]{\boldsymbol{#1}}
\newcommand{\ut}{\tilde{\bs{u}}}
\newcommand{\ebar}{\bar{\bs{\varepsilon}}}

\begin{document}

\title{\parbox{\textwidth}{\raggedright
Ising-Encoded Computational Homogenization: Stress as a Quantum Observable}}

\author{Mourad Halla and Suvranu De}

\affiliation{Department of Mechanical Engineering, Florida A\&M University--Florida State University College of Engineering, Tallahassee, Florida, USA}

\maketitle

\begin{abstract}
Computational homogenization requires repeated solutions of representative
volume element (RVE) problems to determine the effective response transferred
to the macroscale, including the homogenized stress. Here, we develop a
basis-encoded, energy-based variational quantum formulation for linear-elastic
computational homogenization. The reduced finite-element energy of the RVE is
expressed as a quadratic unconstrained binary optimization (QUBO) problem and
mapped to an Ising Hamiltonian, whose ground state represents the equilibrium
fluctuation field on the encoding grid. Using the Hellmann--Feynman theorem,
differentiation of the strain-dependent Hamiltonian with respect to the
prescribed macroscopic strain yields stress observables whose expectation
values recover the conventional volume-averaged homogenized stress. A key
structural result is that, although the equilibrium Hamiltonian is two-local,
the stress observables are one-local. In linear elasticity, the two-qubit
coupling coefficients are independent of macroscopic strain, hence,
differentiation eliminates all two-qubit coupling terms, leaving only identity
and single-qubit Pauli-$Z$ operators. These observables are diagonal in the
same computational basis as the Ising Hamiltonian and can therefore be
evaluated from the same computational-basis measurements used for the energy.
Consequently, the homogenized stress can be obtained directly from the optimized quantum state using the same computational basis measurements as the Ising Hamiltonian, without requiring reconstruction of the microscopic displacement field or an additional measurement basis. Ground states are approximated using the variational quantum eigensolver (VQE), and the formulation is demonstrated on a two-dimensional periodic RVE.
\end{abstract}

\section{Introduction}
\label{sec:intro}

Heterogeneous materials, including fiber and particulate composites,
polycrystalline metals, porous media, and architected solids, exhibit
macroscopic behavior governed by the properties, morphology, and interactions
of their underlying microstructure. Computational homogenization provides a
systematic framework for determining this effective behavior by solving a
boundary-value problem on a representative volume element (RVE) subjected to a
prescribed macroscopic deformation and transferring the resulting response to
the macroscale~\cite{Hill1963,Feyel2000,Geers2010,Yvonnet2019,Andreassen2014}.
In hierarchical multiscale methods such as $\mathrm{FE}^2$
~\cite{Feyel2000,Geers2010}, RVE problems may be solved repeatedly at
macroscopic integration points and over successive load increments, making the
microscale calculations a dominant component of the overall computational
cost. This repeated-solution structure has motivated extensive research on
FFT-based homogenization~\cite{MoulinecSuquet1998,Schneider2021},
reduced-order modeling~\cite{YvonnetHe2007,Hernandez2014},
surrogate approaches~\cite{LeYvonnet2015}, and alternative
multiscale solution strategies~\cite{RahulDe2015a,RahulDe2015b}.

The principal quantity transferred from the RVE is the homogenized stress,
while some macroscopic solution procedures additionally require the effective
tangent. In linear elasticity, the latter reduces to a constant effective
stiffness tensor that can be determined from independent macroscopic strain
states. Conventionally, the effective response is obtained after solving the
microscopic equilibrium problem and evaluating appropriate volume averages of
the microscopic fields. The computational burden therefore lies primarily in
the repeated solution of the RVE problems rather than in the averaging
operation itself, motivating methods that accelerate the microscale solution
or reduce the number and cost of such solves
~\cite{YvonnetHe2007,Hernandez2014,RahulDe2015a,RahulDe2015b}.

Quantum computing provides alternative approaches to the discrete RVE problem.
One route is based on amplitude encoding~\cite{HHL2009,Montanaro2016}, in which
an $n$-dimensional solution vector is represented, after normalization, by the
amplitudes of a quantum state using $O(\log n)$ qubits. Quantum linear-system
algorithms can, under appropriate assumptions on sparsity, conditioning, state
preparation, and required accuracy, offer favorable asymptotic complexity
~\cite{HHL2009,Montanaro2016}. This compact representation, however, introduces
a fundamental readout challenge: measurement does not directly return the
encoded amplitudes, and reconstruction of the complete solution vector can
require a number of measurements that scales with the size of the classical
output and can therefore eliminate the advantages associated with the compact
quantum representation~\cite{Xu2026}. Consequently, for applications in which
only selected quantities of interest are required, it is important to compute
these quantities directly from the quantum state without first reconstructing
the full classical solution.

A different route is based on basis encoding and the variational structure of
the mechanics problem~\cite{Lucas2014,Glover2018,RaisuddinDe2022}. Continuous
mechanical unknowns are represented by finite binary expansions, allowing a
quadratic energy functional to be expressed as a quadratic unconstrained binary
optimization (QUBO) problem and subsequently mapped to an Ising Hamiltonian
~\cite{Lucas2014,Glover2018}. The equilibrium solution is then represented by
the ground state of the encoded Hamiltonian, which may be sought using quantum
annealing or, on gate-based quantum computers, variational algorithms such as
the variational quantum eigensolver (VQE)
~\cite{Peruzzo2014,Kandala2017,Cerezo2021}. Different optimization methods can
be used to minimize the VQE cost function. The optimization can also be
improved using geometry-aware methods based on the quantum natural gradient
and quantum Fisher information
~\cite{Stokes2020,Halla2025QFI,Halla2025CQNG,Halla2025GeoQNG}. Unlike amplitude
encoding where the components of the solution are stored in quantum amplitudes,
basis encoding stores the finite precision representation of the mechanical
unknowns in computational basis bitstrings. This accessibility comes at the
cost of qubit efficiency. For $n$ mechanical degrees of freedom represented by
$L$ bits each, the qubit requirement is $N_b=nL$, rather than $O(\log n)$. The two representations therefore involve fundamentally different tradeoffs between
qubit complexity, solution resolution, and readout.

Energy-based binary formulations have already been explored in computational
mechanics and therefore do not, by themselves, constitute a contribution of
the present work. Ising/QUBO formulations have been used for finite-element
equilibrium and related mechanics problems using quantum annealing and hybrid
quantum--classical methods
~\cite{RaisuddinDe2022,Chang2019,Borle2019}. Nguyen et al.~\cite{Nguyen2024},
for example, formulated nonlinear and history-dependent finite-element problems
as sequences of binary quadratic minimizations within a quantum-annealing-assisted
sequential quadratic programming framework. Variational quantum methods on
gate-based quantum computers have likewise begun to be investigated for
finite-element and computational-mechanics problems~\cite{Liu2024}. The relevant
question for the present work is what structure this representation induces in
the effective response of an RVE.

Quantum algorithms have also recently been developed specifically for
computational homogenization. Liu, Ortiz, and Cirak~\cite{Liu2024} formulated
periodic RVE calculations using the quantum Fourier transform and a fixed-point
iteration related to FFT-based homogenization, establishing a quantum route to
the microscopic RVE problem. This line has recently been extended by Wang,
Ortiz, and Cirak through QAFE$^2$~\cite{Wang2026}, which embeds quantum RVE
calculations within concurrent multiscale finite-element analysis and exploits
quantum superposition to address ensembles of microscopic problems associated
with macroscopic quadrature points. Separately, Xu et al.~\cite{Xu2026}
addressed the readout problem in amplitude-encoded, linear-system-based
computational homogenization and extracted homogenized stress without
reconstructing the complete microscopic solution.

These developments leave a different question largely unexplored: how can the
homogenized stress be represented directly within an energy-based quantum
formulation?

In this work, we investigate this question using a basis-encoded formulation
of linear-elastic computational homogenization. The reduced finite-element RVE
energy is expressed as a QUBO problem and mapped to an Ising Hamiltonian whose
ground state represents the equilibrium fluctuation field on the finite
encoding grid. In classical homogenization, the macroscopic stress follows
from the energetic consistency between the microscopic and macroscopic
problems embodied in the Hill--Mandel condition~\cite{Hill1963} and is
equivalently obtained by volume averaging the microscopic stress. In quantum
mechanics, the Hellmann--Feynman theorem
~\cite{Hellmann1937,Feynman1939} provides the corresponding energy-derivative
framework: for a stationary eigenstate, differentiation of the energy with
respect to a parameter is given by the expectation value of the derivative of
the Hamiltonian. Related energy-derivative formulations of stress are also
well established in quantum mechanics, including the work of Nielsen and
Martin~\cite{NielsenMartin1983}.

Applying this energy-derivative framework to the Ising-encoded RVE
Hamiltonian, differentiation with respect to the prescribed macroscopic strain
yields stress operators whose expectation values recover the conventional
volume-averaged stress of the encoded RVE. This provides direct access to the
homogenized stress from the optimized quantum state, without requiring
recovery of all encoded displacement components. A key result is that these
operators have lower locality than the underlying equilibrium Hamiltonian. In
linear elasticity, the two-qubit interaction coefficients are independent of
macroscopic strain; differentiation therefore eliminates these interactions,
leaving one-local stress observables containing only identity and single-qubit
Pauli-$Z$ terms. Because these observables are diagonal in the same
computational basis as the Ising Hamiltonian, they require no additional
measurement basis.

Against this background, the main contribution of this work is the derivation
of the homogenized stress directly as a quantum observable within an
Ising-encoded variational formulation of linear-elastic computational
homogenization. Although the encoded RVE equilibrium Hamiltonian is two-local,
differentiation with respect to the prescribed macroscopic strain eliminates
the two-qubit interaction terms, leaving one-local stress observables
containing only identity and single-qubit Pauli-$Z$ operators. These
observables recover the conventional volume-averaged homogenized stress and
are diagonal in the same computational basis as the Ising Hamiltonian, so no
additional measurement basis is required.

The remainder of the paper is organized as follows.
Section~\ref{sec:rve} introduces the reduced finite-element RVE problem.
Section~\ref{sec:ising} develops the binary encoding and maps the discrete RVE
energy to the corresponding QUBO and Ising Hamiltonian formulation.
Section~\ref{sec:vqe} briefly reviews the VQE procedure.
Section~\ref{sec:stress} derives the one-local homogenized-stress quantum
observable.
Section~\ref{sec:example} presents the numerical example, followed by
conclusions in Section~\ref{sec:conclusion}.

\section{Reduced RVE equilibrium problem}
\label{sec:rve}
In this section, we briefly review the reduced finite-element RVE problem for
linear elasticity and introduce the notation used in this work. For a more
detailed discussion of computational homogenization and RVE formulations, we
refer to~\cite{Feyel2000,Geers2010,Yvonnet2019,Andreassen2014}.

Let \(\Omega\subset\mathbb{R}^{D}\), with \(D\in\{1,2,3\}\), denote the
representative volume element in the microscopic coordinate \(\bs y\), and
let \(\ebar\) be the prescribed macroscopic strain. The microscopic
displacement is decomposed into an affine contribution and a fluctuation
field,
\begin{equation}
    \bs u(\bs y)
    =
    \ebar\,\bs y
    +
    \ut(\bs y),
    \label{eq:displacement_decomposition}
\end{equation}
where \(\ebar\) acts as the symmetric second-order macroscopic strain tensor
and \(\ut\) is the microscopic displacement fluctuation. The corresponding
microscopic strain is
\begin{equation}
    \bs\varepsilon(\bs y)
    =
    \ebar
    +
    \bs\varepsilon\big(\ut(\bs y)\big).
    \label{eq:strain_decomposition}
\end{equation}
From here on, all strains and stresses are collected in Voigt vector form of
dimension \(\tfrac12 D(D+1)\), with the same symbols \(\ebar\) and
\(\bs\varepsilon\) retained, and \(\bs C(\bs y)\) denotes the corresponding
symmetric elasticity matrix. Under the kinematic RVE constraints considered
here (periodic fluctuations, or fluctuations vanishing on the boundary), the
fluctuation strain averages to zero over \(\Omega\). Consequently, the
prescribed macroscopic strain \(\ebar\) coincides with the volume-averaged
microscopic strain,
\begin{equation}
    \ebar
    =
    \frac{1}{|\Omega|}
    \int_{\Omega}
    \bs\varepsilon(\bs y)\,d\Omega .
\end{equation}

Under the local linear-elastic relation
\begin{equation}
    \bs\sigma(\bs y)
    =
    \bs C(\bs y)\bs\varepsilon(\bs y),
\end{equation}
the volume-averaged strain-energy density is
\begin{equation}
    \Pi(\ut,\ebar)
    =
    \frac{1}{|\Omega|}
    \int_{\Omega}
    \frac{1}{2}
    \bs\varepsilon(\bs y)^{T}
    \bs C(\bs y)
    \bs\varepsilon(\bs y)
    \,d\Omega .
    \label{eq:Pi_cont}
\end{equation}

Throughout, strains and stresses are represented using the engineering-shear
Voigt convention. Thus, for example, the shear strain component is
\(\gamma_{12}=2\varepsilon_{12}\), and
\(\tfrac12\bs\varepsilon^{T}\bs C\bs\varepsilon\) is the strain-energy density.
With this convention, differentiation of the energy density with respect to
\(\bs\varepsilon\) returns the Voigt stress vector without additional factors.

After finite-element discretization, enforcement of the kinematic RVE
constraints, and removal of the rigid-body modes, the independent
fluctuation degrees of freedom are collected in \(\ut\in\mathbb{R}^{n}\).
Let \(\bs B_{e,q}\) denote the reduced strain--displacement matrix at
quadrature point \(q\) of element \(e\). At the quadrature-point level, the
microscopic strain is
\begin{equation}
    \bs\varepsilon_{e,q}
    =
    \ebar
    +
    \bs B_{e,q}\ut .
    \label{eq:discrete_strain}
\end{equation}
With \(w_q\) the quadrature weight and \(\bs J_{e,q}\) the element Jacobian
at that point, the integration weight is
\begin{equation}
    \omega_{e,q}
    =
    w_q\det(\bs J_{e,q}),
    \label{eq:quadrature_weight}
\end{equation}
and the RVE volume is evaluated consistently as
\begin{equation}
    |\Omega|
    =
    \sum_{e,q}\omega_{e,q}.
    \label{eq:rve_volume_quadrature}
\end{equation}

The discretized energy can be written as
\begin{equation}
    \Pi(\ut,\ebar)
    =
    \Pi_0(\ebar)
    -
    \bs f(\ebar)^{T}\ut
    +
    \frac{1}{2}\ut^{T}\bs K\ut ,
    \label{eq:Pi_quad}
\end{equation}
with the fluctuation-independent contribution
\begin{subequations}
\label{eq:Kf}
\begin{equation}
    \Pi_0(\ebar)
    =
    \frac{1}{|\Omega|}
    \sum_{e,q}
    \frac{\omega_{e,q}}{2}
    \ebar^{T}
    \bs C_{e,q}
    \ebar ,
    \label{eq:Pi0}
\end{equation}
the reduced load vector due to the prescribed macroscopic strain,
\begin{equation}
    \bs f(\ebar)
    =
    -
    \frac{1}{|\Omega|}
    \sum_{e,q}
    \omega_{e,q}
    \bs B_{e,q}^{T}
    \bs C_{e,q}
    \ebar ,
    \label{eq:reduced_force}
\end{equation}
and the reduced stiffness matrix
\begin{equation}
    \bs K
    =
    \frac{1}{|\Omega|}
    \sum_{e,q}
    \omega_{e,q}
    \bs B_{e,q}^{T}
    \bs C_{e,q}
    \bs B_{e,q}.
    \label{eq:reduced_stiffness}
\end{equation}
\end{subequations}

Since \(\Pi_0(\ebar)\) does not depend on the fluctuation vector, we minimize the reduced functional
\begin{equation}
    F(\ut)
    =
    \frac{1}{2}\ut^{T}\bs K\ut
    -
    \bs f^{T}\ut .
    \label{eq:reduced_functional}
\end{equation}
Stationarity with respect to \(\ut\) gives the reduced RVE equilibrium
system
\begin{equation}
    \bs K\ut^{*}
    =
    \bs f ,
    \label{eq:reduced}
\end{equation}
where \(\ut^{*}\in\mathbb{R}^{n}\) denotes the equilibrium fluctuation
vector, i.e.\ the unique unconstrained minimizer of the reduced functional \(F\).

We use the following assumption for the reduced linear-elastic system throughout this work.

\begin{assumption}
\label{ass:K}
In exact arithmetic, the reduced stiffness matrix satisfies
\(\bs K=\bs K^{T}\) and \(\bs K\succ0\) after the RVE constraints have been
applied and all rigid-body modes have been removed. Any negligible
antisymmetric component arising solely from numerical round-off may be
removed through
\(\bs K\leftarrow\frac{1}{2}(\bs K+\bs K^{T})\).
Denoting the extreme eigenvalues of \(\bs K\) by
\(0<\lambda_{\min}\le\lambda_{\max}\), its condition number is
\(\kappa=\lambda_{\max}/\lambda_{\min}\). In the linear-elastic setting,
\(\bs K\) is independent of \(\ebar\), whereas \(\bs f(\ebar)\) and
\(\Pi_0(\ebar)\) depend linearly and quadratically on \(\ebar\),
respectively.
\end{assumption}

Positive definiteness follows from the positive definiteness of
\(\bs C(\bs y)\) together with full column rank of the reduced
strain--displacement operator after the kinematic constraints have been
imposed and the rigid-body modes removed.

\section{Ising Hamiltonian formulation}
\label{sec:ising}
We now prepare one of the key ingredients for VQE, namely the Hamiltonian. For this, we map the reduced RVE problem of Section~\ref{sec:rve} to an Ising Hamiltonian through Quadratic Unconstrained Binary Optimization (QUBO) using a fixed-point binary encoding~\cite{Lucas2014,Glover2018}.

We first formulate the QUBO problem. Let \(n\) denote the number of independent reduced fluctuation degrees of freedom introduced in Section~\ref{sec:rve}. Each fluctuation unknown \(\tilde u_j\), with \(j=1,\ldots,n\), is restricted to the interval \([\tilde u_j^{\min},\tilde u_j^{\max}]\) and represented using \(L_j\) binary variables,
\begin{equation}
    \tilde u_j
    =
    \tilde u_j^{\min}
    +
    \Delta_j
    \sum_{\ell=0}^{L_j-1}
    2^\ell b_{j,\ell},
    \qquad
    b_{j,\ell}\in\{0,1\}.
    \label{eq:encoding}
\end{equation}
The spacing between two neighboring encoded values is
\begin{equation}
    \Delta_j
    =
    \frac{
        \tilde u_j^{\max}
        -
        \tilde u_j^{\min}
    }{
        2^{L_j}-1
    }.
    \label{eq:encoding_resolution}
\end{equation}
This gives \(2^{L_j}\) equally spaced values between
\(\tilde u_j^{\min}\) and \(\tilde u_j^{\max}\).

Collecting all bits in \(\bs{b}\in\{0,1\}^{N_b}\), the total number of binary variables is

\begin{equation}
    N_b = \sum_{j=1}^{n} L_j.
    \label{eq:number_binary_variables}
\end{equation}

If the same number of bits \(L\) is used for every unknown, then

\begin{equation}
    N_b = nL.
    \label{eq:number_binary_variables_uniform}
\end{equation}

The encoding of the complete fluctuation vector can be written compactly as
\begin{equation}
    \ut
    =
    \ut^{\min}
    +
    \bs{R}\bs{b},
    \label{eq:encoding_compact}
\end{equation}
where \(\ut^{\min}\in\mathbb{R}^{n}\) collects the lower bounds and
\(\bs{R}\in\mathbb{R}^{n\times N_b}\) is the encoding matrix. The column
of \(\bs{R}\) associated with the bit \(b_{j,\ell}\) contains
\(2^\ell\Delta_j\) in row \(j\) and zeros elsewhere.

Substitution of Eq.~\eqref{eq:encoding_compact} into the reduced functional
\eqref{eq:reduced_functional} gives
\begin{subequations}
\label{eq:Fb}
\begin{equation}
    F(\bs{b})
    =
    C
    +
    \bs{r}^{T}\bs{b}
    +
    \frac{1}{2}
    \bs{b}^{T}\bs{G}\bs{b},
    \label{eq:Fb_function}
\end{equation}
where the constant contribution is
\begin{equation}
    C
    =
    \frac{1}{2}
    (\ut^{\min})^{T}
    \bs{K}
    \ut^{\min}
    -
    \bs{f}^{T}\ut^{\min},
    \label{eq:Fb_constant}
\end{equation}
the linear coefficient vector is
\begin{equation}
    \bs{r}
    =
    \bs{R}^{T}
    \left(
        \bs{K}\ut^{\min}
        -
        \bs{f}
    \right),
    \label{eq:Fb_linear}
\end{equation}
and the quadratic coefficient matrix is
\begin{equation}
    \bs{G}
    =
    \bs{R}^{T}
    \bs{K}
    \bs{R}.
    \label{eq:Fb_quadratic}
\end{equation}
\end{subequations}

Because \(b_a^2=b_a\) for every binary variable, the diagonal terms of
\(\bs{G}\) can be absorbed into the linear coefficients. The resulting
quadratic unconstrained binary optimization (QUBO) problem is
\begin{equation}
    F(\bs{b})
    =
    C
    +
    \sum_{a=1}^{N_b}
    \alpha_a b_a
    +
    \sum_{a<c}
    \beta_{ac}b_ab_c.
    \label{eq:qubo}
\end{equation}
The linear QUBO coefficients are
\begin{equation}
    \alpha_a
    =
    r_a
    +
    \frac{1}{2}G_{aa},
    \qquad
    a=1,\ldots,N_b.
    \label{eq:qubo_linear_coefficient}
\end{equation}
Since \(\bs{G}=\bs{R}^{T}\bs{K}\bs{R}\) inherits the symmetry of
\(\bs{K}\) (Assumption~\ref{ass:K}), the pairwise coefficients are
\begin{equation}
    \beta_{ac}
    =
    G_{ac},
    \qquad
    a<c.
    \label{eq:qubo_quadratic_coefficient}
\end{equation}

We now construct the Ising Hamiltonian from the QUBO in Eq.~\eqref{eq:qubo}
by promoting each binary variable to the operator
\begin{equation}
    \hat b_a
    =
    \frac{I-Z_a}{2},
    \label{eq:binary_pauli_mapping}
\end{equation}
where \(Z_a\) denotes the Pauli-\(Z\) operator acting on qubit \(a\)
\cite{Lucas2014}. Applying this substitution to Eq.~\eqref{eq:qubo}
produces the Ising Hamiltonian
\begin{equation}
    \hat H_F
    =
    C_H I
    +
    \sum_{a=1}^{N_b}
    h_a Z_a
    +
    \sum_{a<c}
    J_{ac}Z_aZ_c.
    \label{eq:HF}
\end{equation}

Here the two-qubit interaction coefficients are
\begin{subequations}
\label{eq:isingcoef}
\begin{equation}
    J_{ac}
    =
    \frac{1}{4}\beta_{ac},
    \qquad
    a<c, 
    \label{eq:ising_J}
\end{equation}
and the single-qubit coefficients are
\begin{equation}
    h_a
    =
    -
    \frac{1}{2}\alpha_a
    -
    \frac{1}{4}
    \left(
        \sum_{c<a}\beta_{ca}
        +
        \sum_{a<c}\beta_{ac}
    \right).
    \label{eq:ising_h}
\end{equation}
The identity contribution is
\begin{equation}
    C_H
    =
    C
    +
    \frac{1}{2}
    \sum_{a=1}^{N_b}\alpha_a
    +
    \frac{1}{4}
    \sum_{a<c}\beta_{ac}.
    \label{eq:ising_constant}
\end{equation}
\end{subequations}

The following theorem shows that the Ising Hamiltonian reproduces the encoded RVE problem exactly.

\begin{theorem}[Exactness of the Hamiltonian encoding]
\label{thm:exact}
The Hamiltonian \(\hat H_F\) is diagonal in the computational basis and
satisfies
\begin{equation}
    \hat H_F
    \lvert\bs{b}\rangle
    =
    F(\bs{b})
    \lvert\bs{b}\rangle
\end{equation}
for every \(\bs{b}\in\{0,1\}^{N_b}\). Consequently, its
computational-basis ground states correspond exactly to the minimizers of
\(F\) over the encoding grid
\begin{equation}
    \mathcal{G}
    =
    \left\{
        \ut^{\min}
        +
        \bs{R}\bs{b}
        \;:\;
        \bs{b}\in\{0,1\}^{N_b}
    \right\}.
    \label{eq:encoding_grid}
\end{equation}
The ground-state energy is therefore
\begin{equation}
    E_0(\hat H_F)
    =
    F_{\min}
    :=
    \min_{\bs{b}\in\{0,1\}^{N_b}}
    F(\bs{b}).
    \label{eq:ground_energy}
\end{equation}
If the continuous reduced solution
\begin{equation}
    \ut^*
    =
    \bs{K}^{-1}\bs{f}
\end{equation}
belongs to \(\mathcal{G}\), every minimizing bitstring decodes to the exact
reduced finite-element solution.
\end{theorem}

\begin{proof}
For every computational-basis state \(\lvert\bs{b}\rangle\), the binary
operator satisfies
\begin{equation}
    \hat b_a
    \lvert\bs{b}\rangle
    =
    b_a
    \lvert\bs{b}\rangle.
\end{equation}
All operators \(\hat b_a\) commute. Replacing each classical binary
variable \(b_a\) in Eq.~\eqref{eq:qubo} by \(\hat b_a\) therefore produces
a diagonal operator whose eigenvalue on \(\lvert\bs{b}\rangle\) is exactly
\(F(\bs{b})\). Expanding the operator products gives
Eqs.~\eqref{eq:HF}--\eqref{eq:isingcoef}.

It follows that the spectrum of \(\hat H_F\) is
\begin{equation}
    \operatorname{spec}(\hat H_F)
    =
    \left\{
        F(\bs{b})
        \;:\;
        \bs{b}\in\{0,1\}^{N_b}
    \right\}.
\end{equation}
Hence, the minimum eigenvalue equals the minimum QUBO value, and the
corresponding computational-basis eigenstates are precisely those associated
with the minimizing bitstrings. Assumption~\ref{ass:K} implies that the
continuous functional has the unique minimizer
\(\ut^*=\bs{K}^{-1}\bs{f}\). If \(\ut^*\in\mathcal{G}\), the minimum over
the encoding grid is attained at this point, and every minimizing bitstring
decodes to \(\ut^*\).
\end{proof}
The error introduced by the finite encoding grid is analyzed in Appendix~\ref{app:error_analysis}.

Before introducing the homogenized-stress observable, we first review the
variational quantum eigensolver (VQE), which provides the variational
framework used to approximate the ground state of the encoded RVE
Hamiltonian.

\section{Variational quantum eigensolver}
\label{sec:vqe}

In this section, we give only a brief overview of the variational quantum
eigensolver (VQE). For a more detailed discussion, we refer the reader to
Ref.~\cite{Tilly2022}.

VQE is a hybrid quantum--classical algorithm that estimates
the lowest eigenvalue of a Hermitian operator $\hat H$ on $N$ qubits
\cite{Peruzzo2014,Cerezo2021}. Writing its spectral decomposition as
$\hat H=\sum_k E_k\lvert E_k\rangle\langle E_k\rvert$ with
$E_0\le E_1\le\cdots$, the target is the ground-state energy $E_0$.
The corresponding Hilbert space has dimension $2^N$, which makes exact
classical calculations difficult as the system size grows.

Using the Rayleigh--Ritz principle, for any normalized state
$\lvert\psi\rangle$,
\begin{equation}
    \langle\psi\rvert\hat H\lvert\psi\rangle
    =
    \sum_k E_k\,\lvert\langle E_k\vert\psi\rangle\rvert^{2}
    \;\ge\;
    E_0,
    \label{eq:vqe_rayleigh}
\end{equation}

Minimizing the left-hand side of Eq.~\eqref{eq:vqe_rayleigh} over a set of
parameterized trial states gives an upper bound on $E_0$. These states are
generated by a parameterized circuit (the \emph{ansatz}) with $p$ trainable
parameters acting on a reference state,
\begin{equation}
    \lvert\psi(\bs\theta)\rangle
    =
    U(\bs\theta)\,\lvert 0\rangle^{\otimes N},
    \qquad
    \bs\theta\in\mathbb{R}^{p},
    \label{eq:vqe_ansatz}
\end{equation}
and one solves the finite-dimensional optimization
\begin{equation}
    \bs\theta^{\star}
    =
    \arg\min_{\bs\theta\in\mathbb{R}^{p}}\,
    E(\bs\theta),
    \qquad
    E(\bs\theta)
    :=
    \langle\psi(\bs\theta)\rvert\hat H\lvert\psi(\bs\theta)\rangle ,
    \label{eq:vqe_objective}
\end{equation}
so that $E(\bs\theta^{\star})\ge E_0$, with equality when the ansatz reaches
the ground space. This defines the hybrid loop: the quantum device
prepares $\lvert\psi(\bs\theta)\rangle$ and estimates $E(\bs\theta)$, while
a classical optimizer updates $\bs\theta$.

To measure the energy on a quantum device, the Hamiltonian is written as
a sum of Pauli operators,
\begin{equation}
    \hat H
    =
    \sum_m c_m\,\hat P_m,
    \qquad
    \hat P_m\in\{I,X,Y,Z\}^{\otimes N},
    \quad
    c_m\in\mathbb{R},
    \label{eq:vqe_pauli}
\end{equation}
so that
\begin{equation}
    E(\bs\theta)
    =
    \sum_m c_m\,
    \langle\psi(\bs\theta)\rvert
    \hat P_m
    \lvert\psi(\bs\theta)\rangle.
    \label{eq:vqe_expansion}
\end{equation}

In the present work, the operator $\hat H$ is the encoded RVE Hamiltonian
$\hat H_F$ introduced in Section~\ref{sec:ising}.

\section{Homogenized stress as a quantum observable}
\label{sec:stress}

We now derive the homogenized stress directly as the expectation value of a
quantum observable within the Ising-encoded RVE formulation. The resulting
stress observable is one-local and diagonal in the same computational basis as
the Hamiltonian.

For this purpose, the strain-dependent term $\Pi_0(\ebar)$, defined in
Eq.~\eqref{eq:Pi0}, must be retained when constructing the full energy
Hamiltonian, giving
\begin{equation}
\hat H_\Pi(\ebar)
=
C_H^{\Pi}(\ebar)\,I
+\sum_a h_a(\ebar)\,Z_a
+\sum_{a<c}J_{ac}\,Z_aZ_c,
\label{eq:HPi}
\end{equation}
which differs from $\hat H_F$ in Eq.~\eqref{eq:HF} only by the shift
$\Pi_0(\ebar)I$ and therefore has the same ground states.

The next lemma gives the relation between the homogenized stress and the
derivative of the energy with respect to the macroscopic strain.

\begin{lemma}[Discrete average-stress identity]
\label{lem:stress}
For every $\ut\in\mathbb{R}^n$ and every Voigt component $\alpha$,
\begin{equation}
\bar{\sigma}_\alpha = \frac{1}{|\Omega|}\sum_{e,q}\omega_{e,q}\,
\bs{E}_\alpha^T\bs{C}_{e,q}\big(\ebar+\bs{B}_{e,q}\ut\big)
=
\frac{\partial \Pi(\ut,\ebar)}{\partial\bar\varepsilon_\alpha}
=
c_\alpha-\bs{p}_\alpha^T\ut,
\label{eq:identity}
\end{equation}
where $\bs{E}_\alpha$ selects component $\alpha$,
$c_\alpha=\partial\Pi_0/\partial\bar\varepsilon_\alpha
=\frac{1}{|\Omega|}\sum_{e,q}\omega_{e,q}\bs{E}_\alpha^T\bs{C}_{e,q}\ebar$,
and
$\bs{p}_\alpha=\partial\bs{f}/\partial\bar\varepsilon_\alpha
=-\frac{1}{|\Omega|}\sum_{e,q}\omega_{e,q}\bs{B}_{e,q}^T\bs{C}_{e,q}
\bs{E}_\alpha$.
\end{lemma}

\begin{proof}
The left-hand side of \eqref{eq:identity} is the discrete volume average of
the microscopic stress component $\alpha$, since
$\bs{C}_{e,q}(\ebar+\bs{B}_{e,q}\ut)$ is the stress at quadrature point
$(e,q)$ and $\bs{E}_\alpha^T$ extracts its $\alpha$-entry. Split it into the
two contributions of the strain decomposition,
\begin{equation}
\frac{1}{|\Omega|}\sum_{e,q}\omega_{e,q}\,
\bs{E}_\alpha^T\bs{C}_{e,q}\ebar
+
\frac{1}{|\Omega|}\sum_{e,q}\omega_{e,q}\,
\bs{E}_\alpha^T\bs{C}_{e,q}\bs{B}_{e,q}\ut .
\label{eq:identity_split}
\end{equation}
The first sum is $c_\alpha$ by definition. In the second, the symmetry
$\bs{C}_{e,q}=\bs{C}_{e,q}^T$ gives
$\bs{E}_\alpha^T\bs{C}_{e,q}\bs{B}_{e,q}
=(\bs{B}_{e,q}^T\bs{C}_{e,q}\bs{E}_\alpha)^T$, so the second sum equals
$-\bs{p}_\alpha^T\ut$ with $\bs{p}_\alpha$ as defined. Thus
\eqref{eq:identity_split} reduces to $c_\alpha-\bs{p}_\alpha^T\ut$.
Independently, differentiating the quadratic energy \eqref{eq:Pi_quad} at
fixed $\ut$, and using that $\bs{K}$ is independent of $\ebar$
(Assumption~\ref{ass:K}), gives
$\partial\Pi/\partial\bar\varepsilon_\alpha
=\partial\Pi_0/\partial\bar\varepsilon_\alpha
-(\partial\bs{f}/\partial\bar\varepsilon_\alpha)^T\ut
=c_\alpha-\bs{p}_\alpha^T\ut$.
This proves Eq.~\eqref{eq:identity}.
\end{proof}

Equation~\eqref{eq:identity} shows that the homogenized stress can be obtained
from the energy derivative. Following the Hellmann--Feynman theorem
~\cite{Hellmann1937,Feynman1939}, we therefore derive the stress operator by
differentiating $\hat H_\Pi$ with respect to the macroscopic strain. Related
energy-derivative formulations of stress are well established in quantum
mechanics, including the work of Nielsen and Martin~\cite{NielsenMartin1983}.
The next theorem gives the resulting operator.

\begin{theorem}[Homogenized-stress observable]\label{thm:observable}
For each Voigt component $\alpha$, define the stress operator
\begin{equation}
\hat{\bar\sigma}_\alpha
:=\frac{\partial\hat H_\Pi(\ebar)}{\partial\bar\varepsilon_\alpha}.
\label{eq:sigma_def}
\end{equation}
Under Assumption~\ref{ass:K} this operator is diagonal in the computational
basis and one-local: with $\bs{t}_\alpha=\bs{R}^T\bs{p}_\alpha$,
\begin{equation}
\hat{\bar\sigma}_\alpha
=A_{\alpha,0}\,I+\sum_{a=1}^{N_b}A_{\alpha,a}\,Z_a,
\qquad
A_{\alpha,0}=c_\alpha-\bs{p}_\alpha^T\ut^{\min}-\tfrac12\sum_a t_{\alpha,a},
\qquad
A_{\alpha,a}=\tfrac12\,t_{\alpha,a}.
\label{eq:sigmaobs}
\end{equation}
In particular, $\hat{\bar\sigma}_\alpha$ contains no two-qubit ($Z_aZ_c$)
terms and commutes with $\hat H_F$.
\end{theorem}

\begin{proof}
By Assumption~\ref{ass:K} the reduced stiffness $\bs{K}$ is independent of
$\ebar$; hence the quadratic coefficient matrix $\bs{G}=\bs{R}^T\bs{K}\bs{R}$
and every Ising coupling $J_{ac}=\tfrac14 G_{ac}$ are also
$\ebar$-independent. Differentiating $\hat H_\Pi$ in \eqref{eq:HPi} with
respect to $\bar\varepsilon_\alpha$ therefore annihilates the entire
two-qubit interaction, and only the identity and single-qubit terms survive,
\begin{equation}
\hat{\bar\sigma}_\alpha
=\frac{\partial C_H^{\Pi}}{\partial\bar\varepsilon_\alpha}\,I
+\sum_a\frac{\partial h_a}{\partial\bar\varepsilon_\alpha}\,Z_a .
\label{eq:sigma_surviving}
\end{equation}
To identify the coefficients, note that $\hat H_\Pi(\ebar)$ is diagonal in
the computational basis for every $\ebar$, so its eigenprojectors are
$\ebar$-independent and the derivative acts on the eigenvalues alone. On a
basis state, $\hat H_\Pi|\bs{b}\rangle
=\Pi(\ut_{\bs{b}},\ebar)|\bs{b}\rangle$ with
$\ut_{\bs{b}}=\ut^{\min}+\bs{R}\bs{b}$, so differentiating and applying
Lemma~\ref{lem:stress} gives
\begin{equation}
\langle\bs{b}|\hat{\bar\sigma}_\alpha|\bs{b}\rangle
=\frac{\partial\Pi(\ut_{\bs{b}},\ebar)}{\partial\bar\varepsilon_\alpha}
=c_\alpha-\bs{p}_\alpha^T\ut_{\bs{b}}
=c_\alpha-\bs{p}_\alpha^T\ut^{\min}-\bs{t}_\alpha^T\bs{b},
\end{equation}
using $\bs{p}_\alpha^T\bs{R}=\bs{t}_\alpha^T$. Thus
$\hat{\bar\sigma}_\alpha=(c_\alpha-\bs{p}_\alpha^T\ut^{\min})\,I
-\sum_a t_{\alpha,a}\hat b_a$ as a diagonal operator; substituting
$\hat b_a=(I-Z_a)/2$ and collecting the identity and $Z_a$ contributions
yields the coefficients \eqref{eq:sigmaobs}. Being a sum of $I$ and $Z_a$
only, $\hat{\bar\sigma}_\alpha$ is diagonal and commutes with the diagonal
$\hat H_F$.
\end{proof}
Theorem~\ref{thm:observable} gives a simple form of the stress operator,
containing only identity and single-qubit $Z$ terms. The next theorem shows
how its expectation value is related to the displacement-based stress
average.

\begin{theorem}[Equivalence and exactness of the stress observable]\label{thm:routes}
Let $|\psi\rangle=\sum_{\bs{b}}c_{\bs{b}}|\bs{b}\rangle$ be an arbitrary
$N_b$-qubit state. Then
\begin{equation}
\langle\psi|\hat{\bar\sigma}_\alpha|\psi\rangle
=\sum_{\bs{b}}|c_{\bs{b}}|^2\,
\bar\sigma_\alpha^{\mathrm{avg}}\!\big(\ut_{\bs{b}}\big),
\qquad
\ut_{\bs{b}}=\ut^{\min}+\bs{R}\bs{b},
\label{eq:routes}
\end{equation}
where $\bar\sigma_\alpha^{\mathrm{avg}}(\cdot)$ is the Gauss-point stress
average. 
\end{theorem}

\begin{proof}
Collect the bits into the diagonal displacement operator
$\hat{\ut}=\ut^{\min}I+\bs{R}\hat{\bs{b}}$, where
$\hat{\bs{b}}=(\hat b_1,\dots,\hat b_{N_b})$ and $\hat b_a=(I-Z_a)/2$. Each
$\hat b_a$ is diagonal with $\hat b_a|\bs{b}\rangle=b_a|\bs{b}\rangle$, so
$\hat{\ut}|\bs{b}\rangle=\ut_{\bs{b}}|\bs{b}\rangle$. From the definitions
\eqref{eq:sigma_def} and Lemma~\ref{lem:stress},
$\hat{\bar\sigma}_\alpha=c_\alpha I-\bs{p}_\alpha^T\hat{\ut}$; substituting
$\hat b_a=(I-Z_a)/2$ and grouping identity and $Z_a$ terms gives the
explicit coefficients \eqref{eq:sigmaobs}. Because every term is diagonal,
\begin{equation}
\langle\psi|\hat{\bar\sigma}_\alpha|\psi\rangle
=\sum_{\bs{b}}|c_{\bs{b}}|^2\,
\langle\bs{b}|\hat{\bar\sigma}_\alpha|\bs{b}\rangle
=\sum_{\bs{b}}|c_{\bs{b}}|^2\big(c_\alpha-\bs{p}_\alpha^T\ut_{\bs{b}}\big),
\end{equation}
and Lemma~\ref{lem:stress} identifies each summand $c_\alpha
-\bs{p}_\alpha^T\ut_{\bs{b}}$ with the Gauss-point average
$\bar\sigma_\alpha^{\mathrm{avg}}(\ut_{\bs{b}})$, proving \eqref{eq:routes}.
A computational basis state has a single nonzero $|c_{\bs{b}}|^2=1$, so the
weighted sum collapses to one term and the two routes coincide exactly.
\end{proof}
For an exact encoded ground state, Eq.~\eqref{eq:routes} identifies the
homogenized stress of the encoded RVE with the expectation value of the stress
operator. The effects of imperfect state preparation and finite-shot
measurements are discussed in Appendix~\ref{app:error_analysis}.

The basis encoding used here requires $N_b=nL$ qubits for $n$ reduced
fluctuation degrees of freedom represented with $L$ bits each. This is less
qubit-efficient than amplitude encoding, which represents an $n$-dimensional
vector using $O(\log n)$ qubits. The present work therefore does not claim an
asymptotic computational speedup over classical or amplitude-encoded
approaches. The relevant advantage for the present formulation is instead the
measurement structure of the homogenized stress: each stress operator contains
only identity and single-qubit Pauli-$Z$ terms and is diagonal in the same
computational basis as the Ising Hamiltonian. Consequently, all stress
components can be estimated from the same computational-basis measurement
samples used to evaluate the Hamiltonian, without requiring an additional
measurement basis.
\section{Numerical example}
\label{sec:example}

In this section, we present a small-scale numerical example. Abaqus~\cite{Abaqus}
is used for the classical FEM solution, while Qiskit~\cite{Qiskit2024} is used
for the quantum simulation.

\begin{figure}[H]
    \centering
    \includegraphics[width=0.65\textwidth]{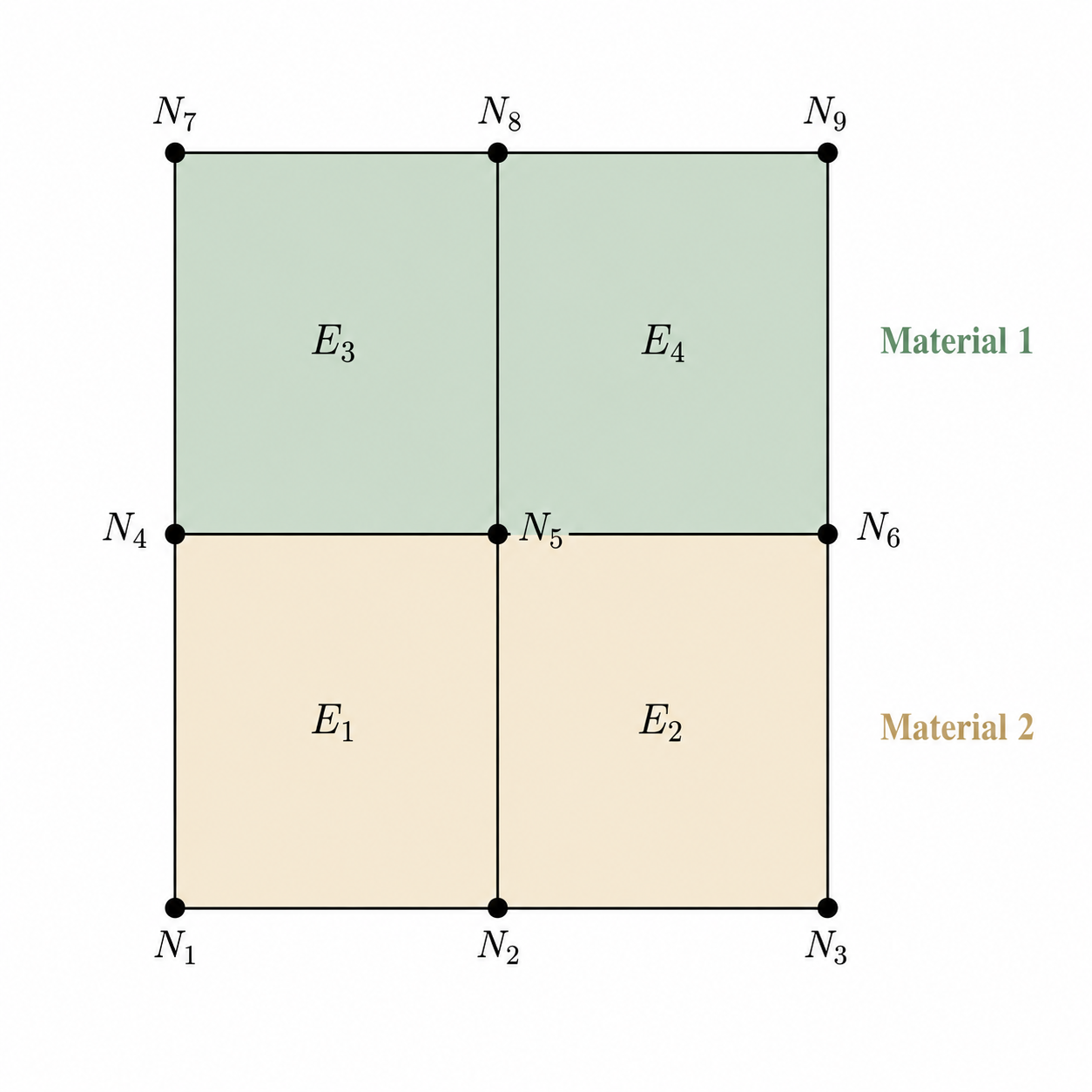}
    \caption{Two-dimensional laminate RVE with four finite elements and nine nodes. The upper and lower layers correspond to Material~1 and Material~2, respectively.}
    \label{fig:2d_rve}
\end{figure}

We consider the two-dimensional periodic laminate RVE shown in
Fig.~\ref{fig:2d_rve}. The RVE consists of two linear-elastic materials and
is discretized using four finite elements with nine nodes. Plane-strain
conditions are assumed. Periodic boundary conditions are imposed on opposite
sides of the RVE. After applying the periodic constraints and removing
rigid-body motion, the system is reduced to six independent fluctuation
degrees of freedom. Additional details of the classical FEM model are given
in Appendix~\ref{app:numerical_details}.

For the VQE calculation, each classical degree of freedom is encoded with
two qubits, giving a 12-qubit Ising Hamiltonian. For this proof-of-concept benchmark,
a common encoding interval of $[-2.8\times10^{-5},\,1.7\times10^{-4}]$ mm is used for all
fluctuation degrees of freedom. With \(L=2\) bits per degree of freedom, this interval
defines a uniform finite-resolution representation of the displacement fluctuations while
keeping the total qubit requirement compact for the present numerical study.

\begin{figure}[H]
  \centering
  \includegraphics[width=0.62\textwidth]{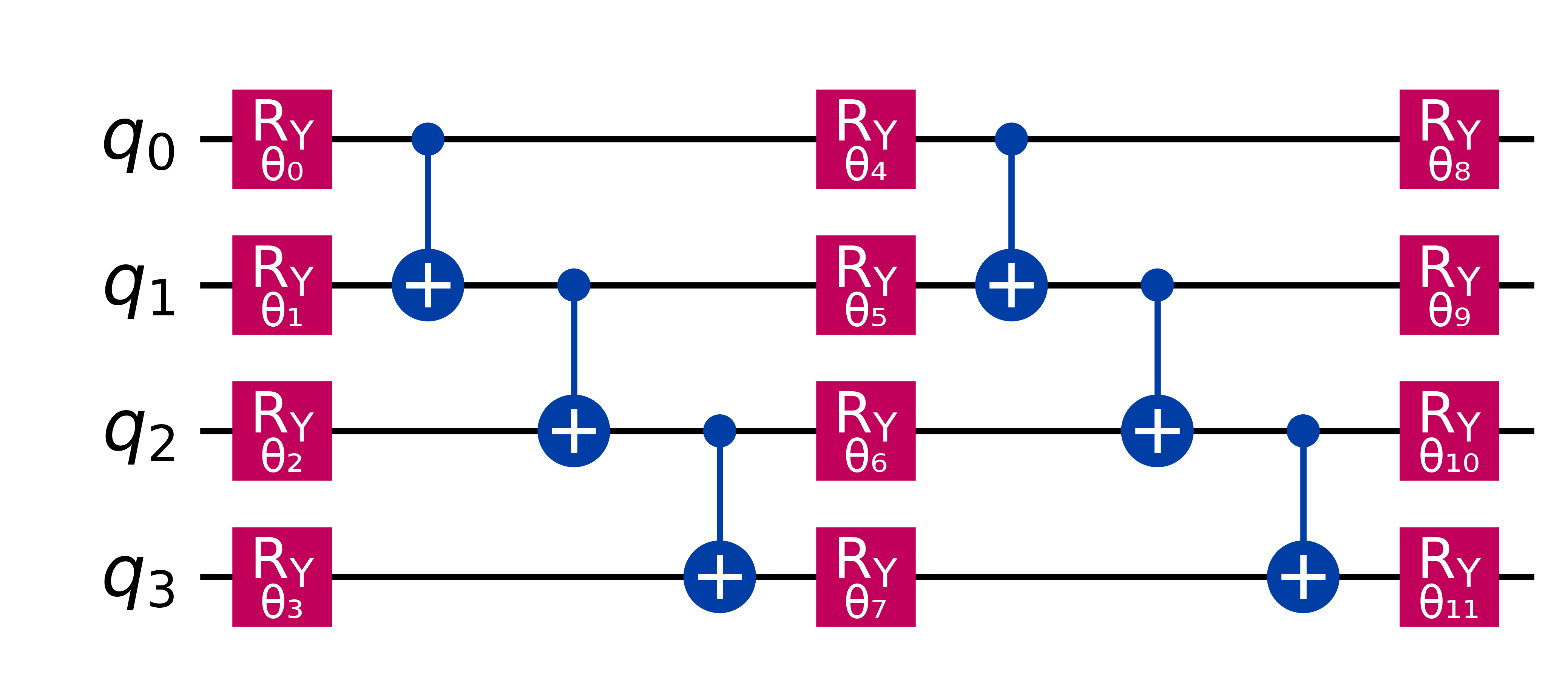}
\caption{
Four-qubit illustration of the $R_y$--CNOT hardware-efficient ansatz used in the VQE calculations. The same circuit structure is extended to the 12-qubit system.
}
  \label{fig:ansatz}
\end{figure}

We use a hardware-efficient ansatz~\cite{Kandala2017} for the VQE optimization,
as illustrated in Fig.~\ref{fig:ansatz}.

\begin{figure}[H]
    \centering
    \includegraphics[width=\linewidth]{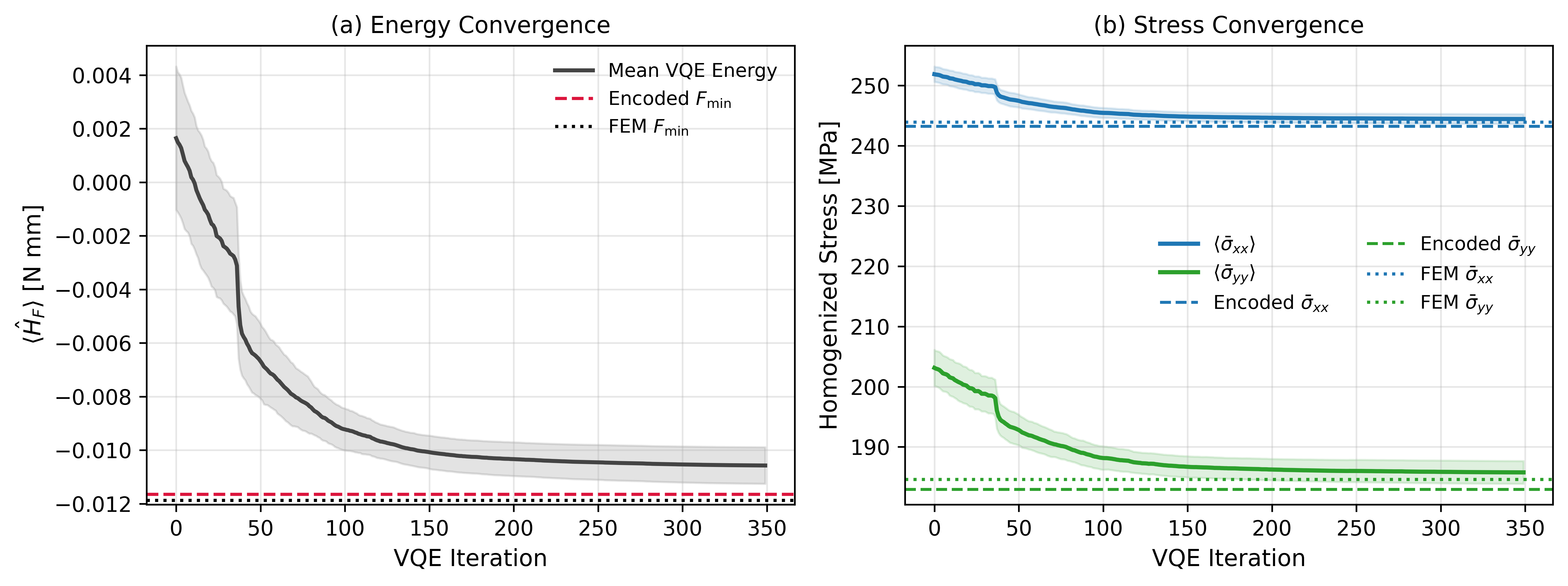}
\caption{
VQE convergence results for the two-dimensional RVE, averaged over 50 runs
initialized from randomly generated parameter points of the VQE ansatz, using
32768 measurement shots per energy evaluation.
(a) Mean VQE energy as a function of iteration. The encoded minimum energy
$F_{\min}$ is shown by the dashed line, while the continuous FEM minimum energy
is shown by the dotted line.
(b) Mean homogenized stress components $\bar{\sigma}_{xx}$ and
$\bar{\sigma}_{yy}$. The encoded stress values are shown by dashed lines, while
the corresponding FEM reference values are shown by dotted lines.
Shaded regions indicate one standard deviation across the 50 VQE runs.
}
    \label{fig:vqe_stress_2d}
\end{figure}

Figure~\ref{fig:vqe_stress_2d} shows the VQE convergence averaged over 50
randomly initialized runs. The VQE energy and the homogenized stress components
converge toward the corresponding encoded and FEM reference values. The energy
exhibits larger deviations than the stress estimates. This behavior is explained
by the different observable structures: the energy Hamiltonian contains both
single-qubit and two-qubit Pauli-$Z$ terms, whereas the stress operators contain
only identity and single-qubit Pauli-$Z$ terms, making the stress estimates less
sensitive to sampling fluctuations.

\begin{figure}[H]
    \centering
    \includegraphics[width=\linewidth]{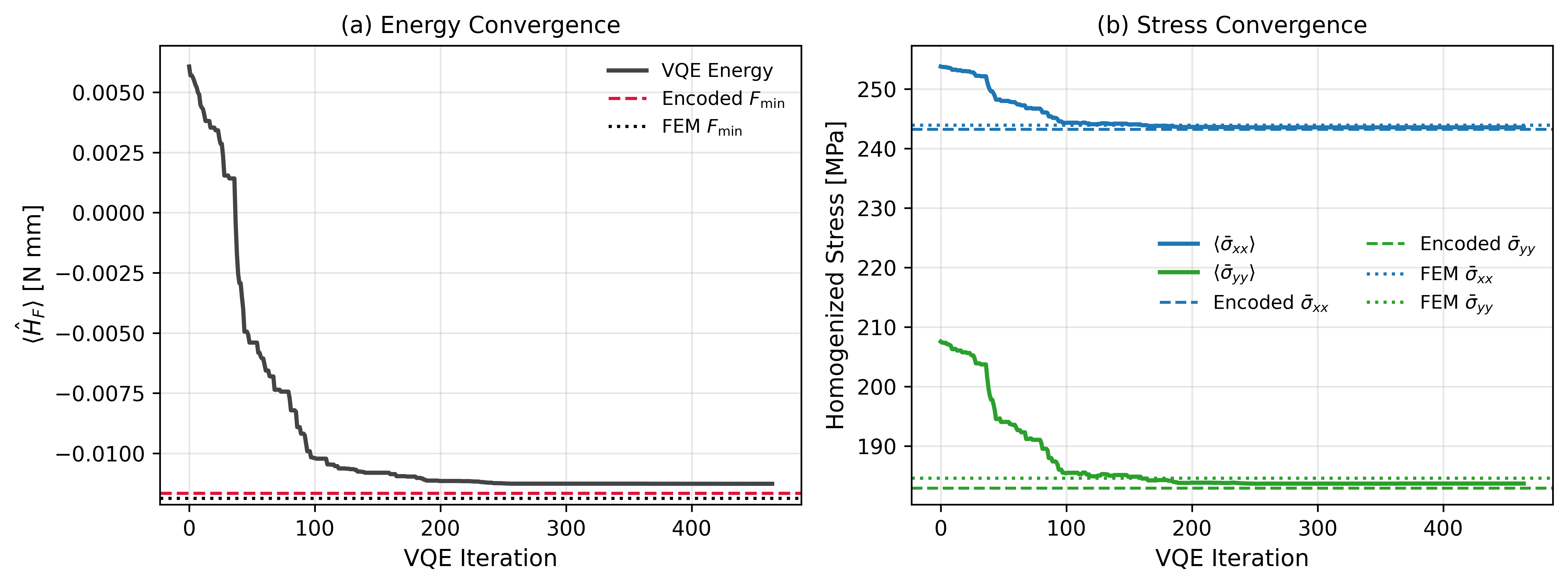}
    \caption{
    VQE convergence for the two-dimensional RVE under the same numerical conditions
    and using the same reference-line conventions as in Fig.~\ref{fig:vqe_stress_2d}.
    The figure shows a run for which the final lowest-energy sampled bitstring satisfies
    $b_{\mathrm{VQE}}=b_{\mathrm{enc}}$.
    }
    \label{fig:best_vqe_stress_2d}
\end{figure}

Figure~\ref{fig:best_vqe_stress_2d} shows a run for which the final
lowest-energy sampled bitstring satisfies
$b_{\mathrm{VQE}}=b_{\mathrm{enc}}$. In this case, we observe smaller
fluctuations in the energy and stress convergence toward the corresponding
encoded reference values.

\begin{figure}[H]
    \centering
    \includegraphics[width=\linewidth]{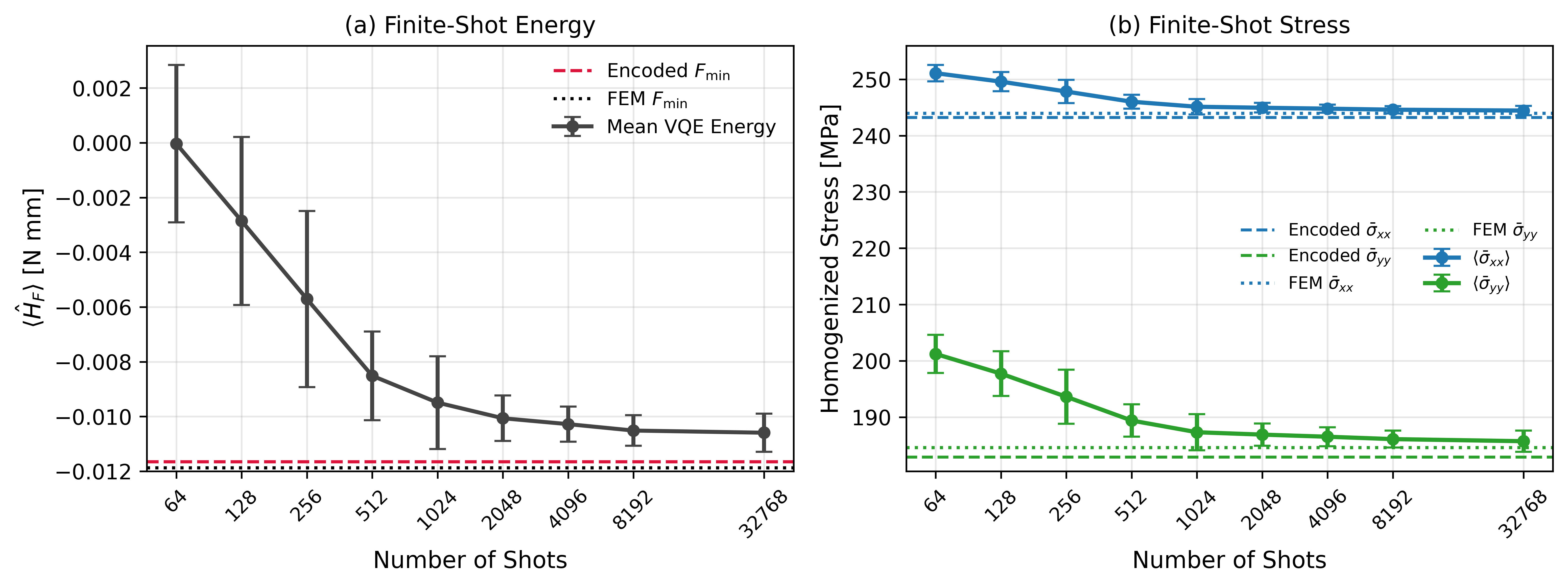}
    \caption{
    Finite-shot VQE results for the two-dimensional RVE. For each shot budget,
    the results are averaged over 50 VQE runs initialized from randomly generated
    parameter points of the VQE ansatz.
    (a) Mean final VQE energy as a function of the number of measurement shots.
    The encoded minimum energy $F_{\min}$ is shown by the dashed line and the
    continuous FEM minimum energy by the dotted line.
    (b) Mean homogenized stress components $\bar{\sigma}_{xx}$ and
    $\bar{\sigma}_{yy}$ as functions of the number of shots. The encoded stress
    values are shown by dashed lines and the corresponding FEM reference values
    by dotted lines. Error bars indicate one standard deviation across the
    50 VQE runs.
    }
    \label{fig:shot_study}
\end{figure}

Figure~\ref{fig:shot_study} shows the finite-shot results, where the averaged
energy and stress estimates are reported as functions of the number of
measurement shots. Increasing the number of shots reduces the statistical
fluctuations and improves the agreement with the corresponding reference values.

\section{Conclusion}
\label{sec:conclusion}

This work developed a basis-encoded, energy-based formulation of
linear-elastic RVE homogenization and established a structural result
concerning the quantum-computed effective response.

Although the encoded equilibrium Hamiltonian is two-local, the homogenized
stress is represented by one-local observables. Differentiation of the
strain-dependent Hamiltonian with respect to the macroscopic strain
eliminates all two-qubit coupling terms, because the reduced stiffness matrix
is strain-independent, and leaves an operator containing only identity and
single-qubit Pauli-$Z$ terms. These observables are diagonal in the same
computational basis as the Hamiltonian and therefore require no additional
measurement basis. Moreover, the homogenized stress can be evaluated directly
from the optimized quantum state without first reconstructing the full encoded
displacement field.

Several practical considerations remain. The attainable solution accuracy
depends on the resolution of the binary encoding. In applications where no
reference solution is available, the encoding intervals must be selected
using a suitable a priori or adaptive strategy. Future work will also explore
amplitude-encoded formulations as a possible route to reduce the qubit
requirements associated with the present basis encoding.

An important extension is material nonlinearity. Under nonlinear material
behavior, the reduced stiffness depends on the macroscopic strain, the
two-qubit couplings may survive differentiation, and the stress observables
are no longer necessarily one-local. Characterizing this structure is an
important direction for future work.

\section*{Acknowledgements}

This work was supported by the U.S. National Science Foundation under
Award No.~2625315, \emph{EAGER: Quantum Algorithms for Reduced-Order Iterative
Methods in Multiscale Solid Mechanics}. The authors gratefully acknowledge
Dr.~Rahul Rahul for helpful discussions on classical multiscale computational
mechanics.

\appendix

\section{Additional details for the numerical example}
\label{app:numerical_details}

\subsection{Classical FEM model}

The RVE is a $1~\mathrm{mm}\times1~\mathrm{mm}$ square consisting of two
linear-elastic layers of equal size. The mesh contains four quadrilateral
finite elements and nine nodes. Plane-strain conditions are assumed. The
material properties and model parameters are given in
Table~\ref{tab:2d_parameters}.

\begin{table}[H]
\centering
\caption{Parameters of the two-dimensional laminate RVE.}
\label{tab:2d_parameters}
\begin{tabular}{lc}
\toprule
Parameter & Value \\
\midrule
RVE dimensions & $1~\mathrm{mm}\times1~\mathrm{mm}$ \\
Material~1 Young's modulus, $E_1$ & $200~\mathrm{GPa}$ \\
Material~1 Poisson's ratio, $\nu_1$ & $0.30$ \\
Material~2 Young's modulus, $E_2$ & $100000~\mathrm{MPa}$ \\
Material~2 Poisson's ratio, $\nu_2$ & $0.30$ \\
Number of elements & $4$ \\
Number of nodes & $9$ \\
Prescribed macroscopic strain, $\bar{\varepsilon}$ &
$[0.001,\;0.0006,\;0]$ \\
\bottomrule
\end{tabular}
\end{table}

Periodic boundary conditions are imposed on opposite sides of the RVE.
After applying the periodic constraints and removing rigid-body motion, six
independent fluctuation degrees of freedom remain.

\subsection{Ising Hamiltonian}
\label{app:ising_2d}

The resulting Ising Hamiltonian used in the VQE simulation is

\begin{small}
\begin{verbatim}
Pauli operators =
['IIIIIIIIIIII', 'IIIIIIIIIIIZ', 'IIIIIIIIIIZI',
 'IIIIIIIIIZII', 'IIIIIIIIZIII', 'IIIIIIIZIIII',
 'IIIIIIZIIIII', 'IIIIIZIIIIII', 'IIIIZIIIIIII',
 'IIIZIIIIIIII', 'IIZIIIIIIIII', 'IZIIIIIIIIII',
 'ZIIIIIIIIIII', 'IIIIIIIIIIZZ', 'IIIIIIIZIIIZ',
 'IIIIIIZIIIIZ', 'IIIZIIIIIIIZ', 'IIZIIIIIIIIZ',
 'IIIIIIIZIIZI', 'IIIIIIZIIIZI', 'IIIZIIIIIIZI',
 'IIZIIIIIIIZI', 'IIIIIIIIZZII', 'IIIIIZIIIZII',
 'IIIIZIIIIZII', 'IZIIIIIIIZII', 'ZIIIIIIIIZII',
 'IIIIIZIIZIII', 'IIIIZIIIZIII', 'IZIIIIIIZIII',
 'ZIIIIIIIZIII', 'IIIIIIZZIIII', 'IIIZIIIZIIII',
 'IIZIIIIZIIII', 'IIIZIIZIIIII', 'IIZIIIZIIIII',
 'IIIIZZIIIIII', 'IZIIIZIIIIII', 'ZIIIIZIIIIII',
 'IZIIZIIIIIII', 'ZIIIZIIIIIII', 'IIZZIIIIIIII',
 'ZZIIIIIIIIII']

Coefficients =
[ 0.0058,  0.0002,  0.0004,  0.0026,  0.0053,
 -0.0004, -0.0008,  0.0026,  0.0051, -0.0003,
 -0.0007, -0.0028, -0.0055,  0.0011, -0.0003,
 -0.0007, -0.0004, -0.0008, -0.0007, -0.0013,
 -0.0008, -0.0016,  0.0011,  0.0002,  0.0004,
 -0.0004, -0.0008,  0.0004,  0.0008, -0.0008,
 -0.0016,  0.0011,  0.0002,  0.0004,  0.0004,
  0.0008,  0.0011, -0.0003, -0.0007, -0.0007,
 -0.0013,  0.0011,  0.0011]
\end{verbatim}
\end{small}


\section{Encoding, stability, and measurement error analysis}
\label{app:error_analysis}
This appendix gives the error and stability results used in the paper. We first discuss the error from the binary encoding, followed by the effects of imperfect state preparation and finite-shot measurements.

\begin{lemma}[Energy--displacement stability]
\label{lem:energy_displacement}
Let Assumption~\ref{ass:K} hold. If \(\ut\in\mathbb{R}^{n}\) satisfies
\(F(\ut)-F(\ut^*)\le\eta\), then
\begin{equation}
    \norm{\ut-\ut^*}_2
    \le
    \sqrt{\frac{2\eta}{\lambda_{\min}}}.
    \label{eq:energy_displacement_bound}
\end{equation}
\end{lemma}

\begin{proof}
Since \(\ut^*\) is the unconstrained minimizer of the differentiable
quadratic \(F\), the first-order optimality condition gives
\(\nabla F(\ut^*)=\bs{0}\). A second-order Taylor expansion of \(F\) about
\(\ut^*\) is exact because \(F\) is quadratic (all derivatives of order
three and higher vanish):
\begin{equation}
    F(\ut)
    =
    F(\ut^*)
    +
    \nabla F(\ut^*)^{T}(\ut-\ut^*)
    +
    \tfrac{1}{2}(\ut-\ut^*)^{T}\bs{K}(\ut-\ut^*),
    \label{eq:lem_taylor}
\end{equation}
where \(\bs{K}=\nabla^2 F\) is the (constant) Hessian. Using
\(\nabla F(\ut^*)=\bs{0}\), the linear term drops and \eqref{eq:lem_taylor}
reduces to the exact identity
\begin{equation}
    F(\ut)-F(\ut^*)
    =
    \tfrac{1}{2}(\ut-\ut^*)^{T}\bs{K}(\ut-\ut^*).
    \label{eq:lem_quadratic_identity}
\end{equation}
By Assumption~\ref{ass:K}, \(\bs{K}\) is symmetric positive definite with
smallest eigenvalue \(\lambda_{\min}>0\), so the Rayleigh quotient is
bounded below,
\begin{equation}
    (\ut-\ut^*)^{T}\bs{K}(\ut-\ut^*)
    \ge
    \lambda_{\min}\,\norm{\ut-\ut^*}_2^{2}.
    \label{eq:lem_rayleigh}
\end{equation}
Combining \eqref{eq:lem_quadratic_identity}, \eqref{eq:lem_rayleigh}, and
the hypothesis \(F(\ut)-F(\ut^*)\le\eta\) yields
\begin{equation}
    \frac{\lambda_{\min}}{2}\,\norm{\ut-\ut^*}_2^{2}
    \le
    \tfrac{1}{2}(\ut-\ut^*)^{T}\bs{K}(\ut-\ut^*)
    =
    F(\ut)-F(\ut^*)
    \le
    \eta.
\end{equation}
Multiplying by \(2/\lambda_{\min}>0\) gives
\(\norm{\ut-\ut^*}_2^{2}\le 2\eta/\lambda_{\min}\), and taking the
nonnegative square root proves
Eq.~\eqref{eq:energy_displacement_bound}.
\end{proof}

We now use this stability result to bound the error introduced by the finite binary encoding.
\begin{theorem}[Encoding error]
\label{thm:encoding}
Let Assumption~\ref{ass:K} hold and suppose each encoding interval contains the
corresponding component of the continuous solution,
\[
    \tilde u_j^{\min}
    \le
    \tilde u_j^*
    \le
    \tilde u_j^{\max},
    \qquad
    j=1,\ldots,n.
\]
Let \(\bs{b}^*\) be any minimizer of \(F\) over \(\{0,1\}^{N_b}\), and define
\[
    \ut_{\mathrm{enc}}
    =
    \ut^{\min}
    +
    \bs{R}\bs{b}^*.
\]
Then
\begin{subequations}
\label{eq:encbound}

\begin{equation}
    F(\ut_{\mathrm{enc}})-F(\ut^*)
    \le
    \frac{\lambda_{\max}}{8}
    \sum_{j=1}^{n}\Delta_j^2,
    \label{eq:encoding_energy_bound}
\end{equation}

and

\begin{equation}
    \norm{\ut_{\mathrm{enc}}-\ut^*}_2
    \le
    \frac{1}{2}
    \left(
    \kappa\sum_{j=1}^{n}\Delta_j^2
    \right)^{1/2}.
    \label{eq:encoding_displacement_bound}
\end{equation}

\end{subequations}
\end{theorem}

\begin{proof}
Let \(\ut_{\mathrm{near}}\in\mathcal{G}\) be a componentwise nearest grid point to \(\ut^*\). Since each \(\tilde u_j^*\) lies within its encoding interval, the nearest representable value in component \(j\) is at most one-half of a grid spacing away. Hence

\begin{equation}
    \bigl|\tilde u_{\mathrm{near},j}-\tilde u_j^*\bigr|
    \le
    \frac{\Delta_j}{2},
    \qquad j=1,\ldots,n,
\end{equation}
and therefore
\begin{equation}
    \norm{\ut_{\mathrm{near}}-\ut^*}_2^{2}
    \le
    \frac{1}{4}\sum_{j=1}^{n}\Delta_j^{2}.
\end{equation}

Using Eq.~\eqref{eq:lem_quadratic_identity}, with
\(\bs e=\ut-\ut^*\), and applying the Rayleigh upper bound gives
\begin{equation}
    F(\ut_{\mathrm{near}})-F(\ut^*)
    =
    \frac{1}{2}\bs e^{T}\bs K\bs e
    \le
    \frac{\lambda_{\max}}{2}\norm{\bs e}_2^2
    \le
    \frac{\lambda_{\max}}{8}
    \sum_{j=1}^{n}\Delta_j^2.
\end{equation}

Since \(\ut_{\mathrm{enc}}\) minimizes \(F\) over \(\mathcal{G}\),
\begin{equation}
    F(\ut_{\mathrm{enc}})-F(\ut^*)
    \le
    F(\ut_{\mathrm{near}})-F(\ut^*)
    \le
    \frac{\lambda_{\max}}{8}\sum_{j=1}^{n}\Delta_j^{2},
\end{equation}
which proves Eq.~\eqref{eq:encoding_energy_bound}.

Finally, applying Lemma~\ref{lem:energy_displacement} with
\(\eta=\tfrac{\lambda_{\max}}{8}\sum_j\Delta_j^{2}\) gives
\begin{equation}
    \norm{\ut_{\mathrm{enc}}-\ut^*}_2
    \le
    \frac{1}{2}
    \left(\kappa\sum_{j=1}^{n}\Delta_j^{2}\right)^{1/2},
\end{equation}
which proves Eq.~\eqref{eq:encoding_displacement_bound}.
\end{proof}

As a direct consequence of the displacement bound, we also obtain an estimate for the equilibrium residual.
\begin{corollary}[Equilibrium residual]
\label{cor:residual}
Under the hypotheses of Theorem~\ref{thm:encoding},
\begin{equation}
    \norm{\bs{K}\ut_{\mathrm{enc}}-\bs{f}}_2
    \le
    \lambda_{\max}\norm{\ut_{\mathrm{enc}}-\ut^*}_2
    \le
    \frac{\lambda_{\max}}{2}\left(\kappa\sum_{j=1}^{n}\Delta_j^2\right)^{1/2}.
    \label{eq:encoding_residual_bound}
\end{equation}
\end{corollary}

\begin{proof}
Since \(\bs{K}\ut^*=\bs{f}\),
\begin{equation}
    \bs{K}\ut_{\mathrm{enc}}-\bs{f}
    =
    \bs{K}(\ut_{\mathrm{enc}}-\ut^*).
\end{equation}
Therefore,
\begin{equation}
    \norm{\bs{K}\ut_{\mathrm{enc}}-\bs{f}}_2
    \le
    \norm{\bs{K}}_2\norm{\ut_{\mathrm{enc}}-\ut^*}_2
    =
    \lambda_{\max}\norm{\ut_{\mathrm{enc}}-\ut^*}_2.
\end{equation}
The second inequality follows directly from
Eq.~\eqref{eq:encoding_displacement_bound}.
\end{proof}
The next result considers the effect of imperfect state preparation and
finite-shot measurements on the stress observable.

\begin{proposition}[Stress-observable stability under state-preparation and shot noise]\label{prop:state}
Let the encoded ground state $|\psi_0\rangle$ be nondegenerate with spectral
gap $\gamma_g>0$, and let $|\psi\rangle$ have energy error
$\eta=\langle\psi|\hat H_F|\psi\rangle-F_{\min}\ge 0$. Then
\begin{equation}
1-|\langle\psi_0|\psi\rangle|^2\le\frac{\eta}{\gamma_g},
\qquad
\big|\langle\hat{\bar\sigma}_\alpha\rangle_\psi
-\langle\hat{\bar\sigma}_\alpha\rangle_{\psi_0}\big|
\le 2\,\norm{\hat{\bar\sigma}_\alpha}\sqrt{\frac{\eta}{\gamma_g}},
\label{eq:stability}
\end{equation}
with $\norm{\hat{\bar\sigma}_\alpha}\le|A_{\alpha,0}|+\sum_a|A_{\alpha,a}|$;
estimating $\langle\hat{\bar\sigma}_\alpha\rangle$ to precision
$\epsilon_\sigma$ requires
$N_{\mathrm{shots}}=\mathcal{O}\!\big(\mathrm{Var}(\hat{\bar\sigma}_\alpha)
/\epsilon_\sigma^2\big)$.
\end{proposition}

\begin{proof}
Expand the prepared state in the eigenbasis of $\hat H_F$,
$|\psi\rangle=\sum_{k\ge0}a_k|E_k\rangle$, with
$E_0=F_{\min}$ and $E_k\ge F_{\min}+\gamma_g$ for $k\ge1$.
Writing $p_0=|\langle\psi_0|\psi\rangle|^2=|a_0|^2$, we have
\begin{equation}
F_{\min}+\eta
=\sum_{k\ge0}|a_k|^2E_k
\ge
p_0F_{\min}+(1-p_0)(F_{\min}+\gamma_g).
\end{equation}
Therefore,
\begin{equation}
1-p_0
\le
\frac{\eta}{\gamma_g}.
\end{equation}

Let $\rho=|\psi\rangle\langle\psi|$ and
$\rho_0=|\psi_0\rangle\langle\psi_0|$. For two pure states,
\begin{equation}
\norm{\rho-\rho_0}_1
=
2\sqrt{1-p_0}.
\end{equation}
Using the matrix H\"older inequality,
\begin{align}
\big|\langle\hat{\bar\sigma}_\alpha\rangle_\psi
-\langle\hat{\bar\sigma}_\alpha\rangle_{\psi_0}\big|
&=
\big|\operatorname{tr}
[\hat{\bar\sigma}_\alpha(\rho-\rho_0)]\big| \nonumber\\
&\le
\norm{\hat{\bar\sigma}_\alpha}\,
\norm{\rho-\rho_0}_1 \nonumber\\
&\le
2\norm{\hat{\bar\sigma}_\alpha}
\sqrt{\frac{\eta}{\gamma_g}}.
\end{align}

From Theorem~\ref{thm:observable},
\[
\hat{\bar\sigma}_\alpha
=
A_{\alpha,0}I+\sum_{a=1}^{N_b}A_{\alpha,a}Z_a.
\]
Since $\norm{I}=\norm{Z_a}=1$, the triangle inequality gives
\begin{equation}
\norm{\hat{\bar\sigma}_\alpha}
\le
|A_{\alpha,0}|
+\sum_{a=1}^{N_b}|A_{\alpha,a}|.
\end{equation}

Finally, because $\hat{\bar\sigma}_\alpha$ is diagonal, computational-basis
measurements give an unbiased estimator of
$\langle\hat{\bar\sigma}_\alpha\rangle_\psi$. Averaging
$N_{\mathrm{shots}}$ independent samples gives variance
\begin{equation}
\frac{\mathrm{Var}_\psi(\hat{\bar\sigma}_\alpha)}
{N_{\mathrm{shots}}}.
\end{equation}
Hence,
\begin{equation}
N_{\mathrm{shots}}
=
\mathcal{O}\!\left(
\frac{\mathrm{Var}_\psi(\hat{\bar\sigma}_\alpha)}
{\epsilon_\sigma^2}
\right).
\end{equation}
Since $\hat{\bar\sigma}_\alpha$ and $\hat H_F$ are diagonal in the same
basis, the same computational-basis measurements used for the energy can
also be used for the stress.
\end{proof}


\newpage

\begin{thebibliography}{99}

\bibitem{Feyel2000}
F. Feyel and J.-L. Chaboche,
``FE\textsuperscript{2} multiscale approach for modelling the elastoviscoplastic behaviour of long fibre SiC/Ti composite materials,''
\emph{Comput. Methods Appl. Mech. Eng.} \textbf{183}, 309--330 (2000).
\url{https://doi.org/10.1016/S0045-7825(99)00224-8}

\bibitem{Geers2010}
M. G. D. Geers, V. G. Kouznetsova, and W. A. M. Brekelmans,
``Multi-scale computational homogenization: Trends and challenges,''
\emph{J. Comput. Appl. Math.} \textbf{234}, 2175--2182 (2010).
\url{https://doi.org/10.1016/j.cam.2009.08.077}

\bibitem{Yvonnet2019}
J. Yvonnet,
\emph{Computational Homogenization of Heterogeneous Materials with Finite Elements},
Solid Mechanics and Its Applications, Vol. 258, Springer (2019).
\url{https://doi.org/10.1007/978-3-030-18383-7}

\bibitem{Andreassen2014}
E. Andreassen and C. S. Andreasen,
``How to determine composite material properties using numerical homogenization,''
\emph{Comput. Mater. Sci.} \textbf{83}, 488--495 (2014).
\url{https://doi.org/10.1016/j.commatsci.2013.09.006}


\bibitem{Hill1963}
R. Hill,
``Elastic properties of reinforced solids: Some theoretical principles,''
\emph{J. Mech. Phys. Solids} \textbf{11}, 357--372 (1963).
\url{https://doi.org/10.1016/0022-5096(63)90036-X}


\bibitem{MoulinecSuquet1998}
H. Moulinec and P. Suquet,
``A numerical method for computing the overall response of nonlinear composites with complex microstructure,''
\emph{Comput. Methods Appl. Mech. Eng.} \textbf{157}, 69--94 (1998).
\url{https://doi.org/10.1016/S0045-7825(97)00218-1}


\bibitem{Schneider2021}
M. Schneider,
``A review of nonlinear FFT-based computational homogenization methods,''
\emph{Acta Mech.} \textbf{232}, 2051--2100 (2021).
\url{https://doi.org/10.1007/s00707-021-02962-1}


\bibitem{YvonnetHe2007}
J. Yvonnet and Q.-C. He,
``The reduced model multiscale method (R3M) for the non-linear homogenization of hyperelastic media at finite strains,''
\emph{J. Comput. Phys.} \textbf{223}, 341--368 (2007).
\url{https://doi.org/10.1016/j.jcp.2006.09.019}


\bibitem{Hernandez2014}
J. A. Hern\'andez, J. Oliver, A. E. Huespe, M. A. Caicedo, and J. C. Cante,
``High-performance model reduction techniques in computational multiscale homogenization,''
\emph{Comput. Methods Appl. Mech. Eng.} \textbf{276}, 149--189 (2014).
\url{https://doi.org/10.1016/j.cma.2014.03.011}


\bibitem{LeYvonnet2015}
B. A. Le and J. Yvonnet,
``Computational homogenization of nonlinear elastic materials using neural networks,''
\emph{Int. J. Numer. Methods Eng.} \textbf{104}, 1061--1084 (2015).
\url{https://doi.org/10.1002/nme.4953}


\bibitem{RahulDe2015a}
Rahul and S. De,
``Multiscale modeling of polycrystalline materials with Jacobian-free multiscale method (JFMM),''
\emph{Comput. Mech.} \textbf{55}, 643--657 (2015).
\url{https://doi.org/10.1007/s00466-015-1128-3}


\bibitem{RahulDe2015b}
Rahul and S. De,
``Analysis of the Jacobian-free multiscale method (JFMM),''
\emph{Comput. Mech.} \textbf{56}, 769--783 (2015).
\url{https://doi.org/10.1007/s00466-015-1200-z}

\bibitem{Nguyen2024}
V.-D. Nguyen, L. Wu, F. Remacle, and L. Noels,
``A quantum annealing-sequential quadratic programming assisted finite element
simulation for non-linear and history-dependent mechanical problems,''
\emph{Eur. J. Mech. A/Solids} \textbf{105}, 105254 (2024).
\url{https://doi.org/10.1016/j.euromechsol.2024.105254}

\bibitem{Wang2026}
Y. Wang, M. Ortiz, and F. Cirak,
``QAFE$^2$: Quantum accelerated multiscale finite element analysis,''
\emph{Comput. Methods Appl. Mech. Eng.} \textbf{461}, 119204 (2026).
\url{https://doi.org/10.1016/j.cma.2026.119204}

\bibitem{NielsenMartin1983}
O. H. Nielsen and R. M. Martin,
``First-principles calculation of stress,''
\emph{Phys. Rev. Lett.} \textbf{50}, 697--700 (1983).
\url{https://doi.org/10.1103/PhysRevLett.50.697}



\bibitem{HHL2009}
A. W. Harrow, A. Hassidim, and S. Lloyd,
``Quantum algorithm for linear systems of equations,''
\emph{Phys. Rev. Lett.} \textbf{103}, 150502 (2009).
\url{https://doi.org/10.1103/PhysRevLett.103.150502}

\bibitem{Montanaro2016}
A. Montanaro and S. Pallister,
``Quantum algorithms and the finite element method,''
\emph{Phys. Rev. A} \textbf{93}, 032324 (2016).
\url{https://doi.org/10.1103/PhysRevA.93.032324}

\bibitem{Lucas2014}
A. Lucas,
``Ising formulations of many NP problems,''
\emph{Front. Phys.} \textbf{2}, 5 (2014).
\url{https://doi.org/10.3389/fphy.2014.00005}

\bibitem{Glover2018}
F. Glover, G. Kochenberger, and Y. Du,
``A tutorial on formulating and using QUBO models,''
arXiv:1811.11538 (2018).
\url{https://doi.org/10.48550/arXiv.1811.11538}

\bibitem{Farhi2014}
E. Farhi, J. Goldstone, and S. Gutmann,
``A quantum approximate optimization algorithm,''
arXiv:1411.4028 (2014).
\url{https://doi.org/10.48550/arXiv.1411.4028}

\bibitem{Tilly2022}
J. Tilly et al.,
``The variational quantum eigensolver: A review of methods and best practices,''
\emph{Phys. Rep.} \textbf{986}, 1--128 (2022).
\url{https://doi.org/10.1016/j.physrep.2022.08.003}

\bibitem{Peruzzo2014}
A. Peruzzo, J. McClean, P. Shadbolt, M.-H. Yung, X.-Q. Zhou, P. J. Love, A. Aspuru-Guzik, and J. L. O'Brien,
``A variational eigenvalue solver on a photonic quantum processor,''
\emph{Nat. Commun.} \textbf{5}, 4213 (2014).
\url{https://doi.org/10.1038/ncomms5213}

\bibitem{Kandala2017}
A. Kandala, A. Mezzacapo, K. Temme, M. Takita, M. Brink, J. M. Chow, and J. M. Gambetta,
``Hardware-efficient variational quantum eigensolver for small molecules and quantum magnets,''
\emph{Nature} \textbf{549}, 242--246 (2017).
\url{https://doi.org/10.1038/nature23879}

\bibitem{Cerezo2021}
M. Cerezo, A. Arrasmith, R. Babbush, S. C. Benjamin, S. Endo, K. Fujii, J. R. McClean, K. Mitarai, X. Yuan, L. Cincio, and P. J. Coles,
``Variational quantum algorithms,''
\emph{Nat. Rev. Phys.} \textbf{3}, 625--644 (2021).
\url{https://doi.org/10.1038/s42254-021-00348-9}

\bibitem{Preskill2018}
J. Preskill,
``Quantum computing in the NISQ era and beyond,''
\emph{Quantum} \textbf{2}, 79 (2018).
\url{https://doi.org/10.22331/q-2018-08-06-79}

\bibitem{Stokes2020}
J. Stokes, J. Izaac, N. Killoran, and G. Carleo,
``Quantum natural gradient,''
\emph{Quantum} \textbf{4}, 269 (2020).
\url{https://doi.org/10.22331/q-2020-05-25-269}

\bibitem{Halla2025QFI}
M. Halla,
``Estimation of quantum Fisher information via Stein's identity in variational quantum algorithms,''
\emph{Quantum} \textbf{9}, 1798 (2025).
\url{https://doi.org/10.22331/q-2025-07-21-1798}

\bibitem{Halla2025CQNG}
M. Halla,
``Modified conjugate quantum natural gradient,''
\emph{EPJ Quantum Technol.} \textbf{12}, 123 (2025).
\url{https://doi.org/10.1140/epjqt/s40507-025-00432-4}

\bibitem{Halla2025GeoQNG}
M. Halla,
``Quantum natural gradient with geodesic corrections for small shallow quantum circuits,''
\emph{Phys. Scr.} \textbf{100}, 055121 (2025).
\url{https://doi.org/10.1088/1402-4896/add05e}

\bibitem{RaisuddinDe2022}
O. M. Raisuddin and S. De,
``FEqa: Finite element computations on quantum annealers,''
\emph{Comput. Methods Appl. Mech. Eng.} \textbf{395}, 115014 (2022).
\url{https://doi.org/10.1016/j.cma.2022.115014}

\bibitem{Chang2019}
C. C. Chang, A. Gambhir, T. S. Humble, and S. Sota,
``Quantum annealing for systems of polynomial equations,''
\emph{Sci. Rep.} \textbf{9}, 10258 (2019).
\url{https://doi.org/10.1038/s41598-019-46729-0}

\bibitem{Borle2019}
A. Borle and S. J. Lomonaco,
``Analyzing the quantum annealing approach for solving linear least squares problems,''
in \emph{WALCOM: Algorithms and Computation}, LNCS 11355, 289--301, Springer (2019).
\url{https://doi.org/10.1007/978-3-030-10564-8_23}

\bibitem{Liu2024}
B. Liu, M. Ortiz, and F. Cirak,
``Towards quantum computational mechanics,''
\emph{Comput. Methods Appl. Mech. Eng.} \textbf{432}, 117403 (2024).
\url{https://doi.org/10.1016/j.cma.2024.117403}

\bibitem{Xu2026}
Y. Xu, Z. Kuang, L. Li, Q. Huang, J. Yang, K. Huang, Z. Wang, H. Fan, and H. Hu,
``Towards quantum computing enhanced general computational homogenization without readout problem,''
\emph{Int. J. Mech. Sci.}, art. 111214 (2026).
\url{https://doi.org/10.1016/j.ijmecsci.2026.111214}

\bibitem{Hellmann1937}
H. Hellmann,
\emph{Einf\"uhrung in die Quantenchemie},
Franz Deuticke, Leipzig (1937).

\bibitem{Feynman1939}
R. P. Feynman,
``Forces in molecules,''
\emph{Phys. Rev.} \textbf{56}, 340--343 (1939).
\url{https://doi.org/10.1103/PhysRev.56.340}


\bibitem{Abaqus}
Dassault Syst\`emes,
``Abaqus 2024 Documentation,''
SIMULIA, Dassault Syst\`emes (2024).

\bibitem{Qiskit2024}
A. Javadi-Abhari, M. Treinish, K. Krsulich, C. J. Wood, J. Lishman, J. Gacon,
S. Martiel, P. D. Nation, L. S. Bishop, A. W. Cross, B. R. Johnson, and
J. M. Gambetta,
``Quantum computing with Qiskit,''
arXiv:2405.08810 (2024).
\url{https://doi.org/10.48550/arXiv.2405.08810}

\end{thebibliography}
\end{document}